\documentclass[prl,twocolumn,showpacs,amsmath,amssymb]{revtex4-1}
\usepackage{amssymb}
\usepackage{graphicx}
\usepackage{bm}
\usepackage{amsmath}

\newcommand{\beq}{\begin{equation}}
\newcommand{\eeq}{\end{equation}}
\newcommand{\beqn}{\begin{eqnarray}}
\newcommand{\eeqn}{\end{eqnarray}}

\renewcommand{\vec}[1]{\mbox{\boldmath$#1$}}
\begin{document}

\title{Long-Range Spin-Triplet Helix in Proximity Induced Superconductivity in Spin-Orbit-Coupled Systems}

\author{Xin Liu, J. K. Jain, and Chao-Xing Liu}
\affiliation{Department of Physics, The Pennsylvania State University, University Park,
Pennsylvania 16802-6300}
\date{\today}

\begin{abstract}
We study proximity induced triplet superconductivity in a spin-orbit-coupled system, and show that the \textbf{d} vector of the induced triplet superconductivity undergoes precession that can be controlled by varying the relative strengths of Rashba and Dresselhaus spin-orbit couplings.
In particular, a long-range spin-triplet helix is predicted when these two spin-orbit couplings have equal strengths. We also study the Josephson junction geometry and show that a transition between 0 and $\pi$ junctions can be induced by controlling the spin-orbit coupling with a gate voltage. An experimental setup is proposed to verify these effects. Conversely, the observation of these effects can serve as a direct confirmation of triplet superconductivity.
\end{abstract}

 \maketitle

{\it Introduction -}
Crucial to the success of spintronics \cite{Zutic:2004_a} are injection of spin, its long decay length and its manipulation. The study of spin transport in a superconductor has given rise to the subfield known as superconducting spintronics \cite{Ohnishi2010,Quay2013,Wakamura2014}. One may wonder if the spin-1 of Cooper pairs in a triplet superconductor can play a similar role as the electron spin in spintronics. The observation of surprisingly long-range proximity effect in a superconductor (SC)/ferromagnet (FM) junction \cite{Keizer2006a,Wang2010a,Robinson2010a,Khaire2010a,Klose2012a,Robinson2012a,Leksin2012} has been interpreted in terms of an injection into the FM of triplet Cooper pairs with a long decay length \cite{Bergeret2001a,Eschrig2008a,Bergeret2005a,Buzdin2005a,Takei2012a,Bergeret2013a}. However, it is unclear how to manipulate the long-range part of the induced triplet pair. 

We propose here a geometry in which the triplet pairs are injected into a material with spin-orbit coupling (SOC) and show, theoretically, that they can be manipulated by varying the relative strengths of the Rashba and Dresselhaus SOCs. In particular, we predict a long-range spin-triplet helix, which can be verified by observing a $0-\pi$ transition in Josephson junctions as a function of the SOC strengths. We show that the effect is robust against any spin independent scattering. Proximity effect in SOC materials has been considered previously,\cite{Yang2009,Yang2010} but with only Rashba SOC, which does not produce long-range effects discussed below.

\begin{figure}[ht]
\centering
\begin{tabular}{l}
\includegraphics[width=0.75\columnwidth]{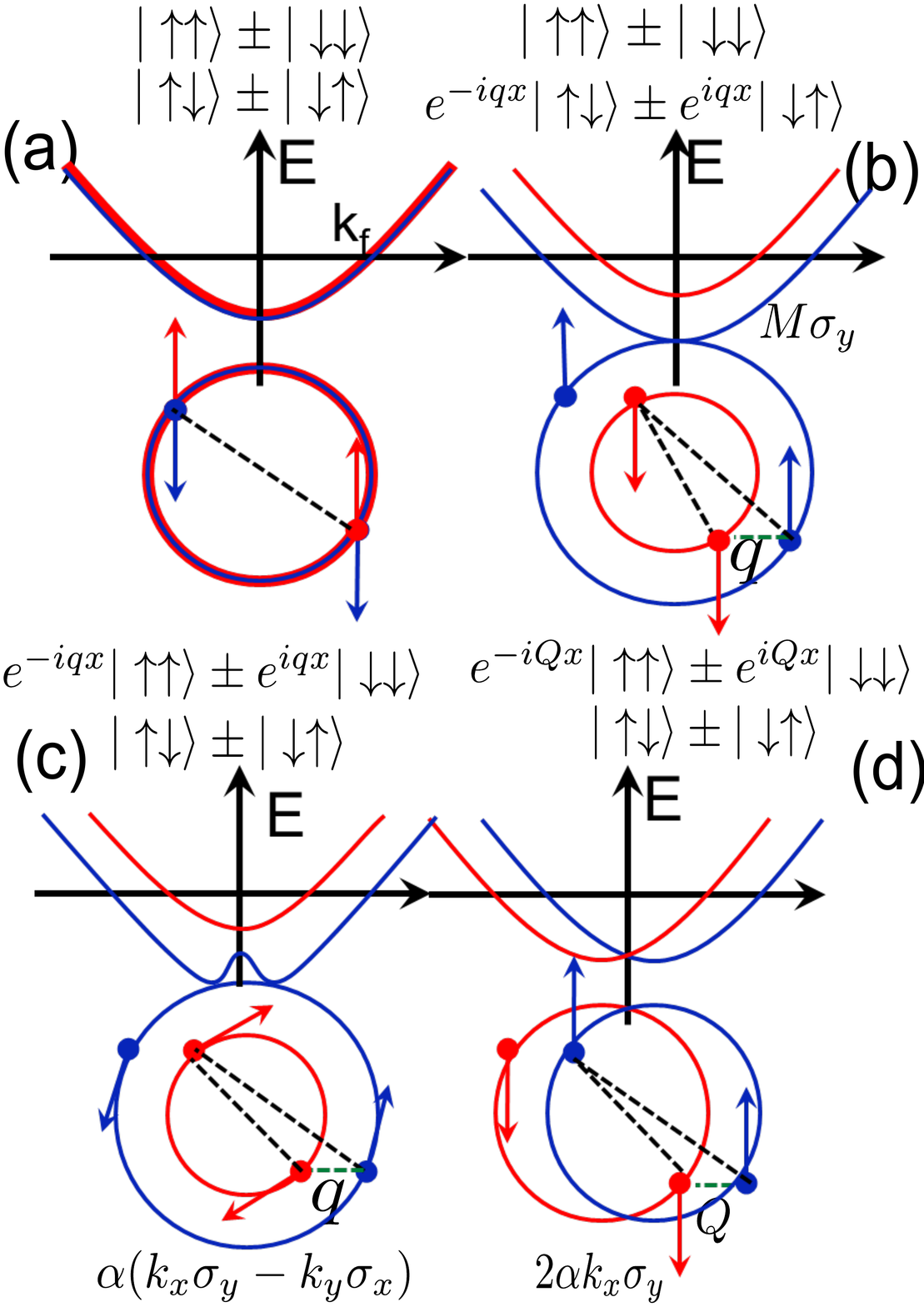}
\end{tabular}
\caption{Energy dispersion and Fermi surfaces are shown for (a) normal metals, (b) ferromagnets, (c) a 2DEG with Rashba SOC and (d) a 2DEG with equal strengths of Rashba and Dresselhaus SOCs. The possible forms of spin states of Cooper pairs, including singlet and triplet pairs, are also illustrated in the figures. For $k_y\neq 0,k_x=0$, the gap between two spin bands in (c) is $|2\alpha k_y|$.}
\label{pairing}
\end{figure}

Before presenting the detailed microscopic theory, we first illustrate the underlying physics, shown in Fig~\ref{pairing}.  In the absence of magnetization and SOC, four kinds of Cooper pairs (singlet $|\uparrow \downarrow \rangle-|\downarrow\uparrow \rangle$ and triplet pairs $|\uparrow \downarrow \rangle+|\downarrow\uparrow \rangle$, $|\uparrow\uparrow\rangle \pm |\downarrow\downarrow \rangle$) are allowed with a zero center-of-mass momentum. The magnetization breaks the degeneracy between $|k,\uparrow\rangle$ and $|-k,\downarrow\rangle$. It will lead to a spatially modulated oscillation $e^{-iqx}|\uparrow\downarrow\rangle\pm e^{iqx}|\downarrow\uparrow\rangle$ \cite{Demler1997,Eschrig2011} for the Cooper pairs with opposite spins but leave the pairs  $|\uparrow\uparrow\rangle \pm |\downarrow\downarrow \rangle$ unchanged, as shown in Fig~\ref{pairing}(b). (Here we assume that the system is uniform along y and z directions so that the center-of-mass momentum of pairs is always zero along these directions.) 
On the contrary, SOC breaks the degeneracy between $|k,\uparrow(\downarrow)\rangle$ and $|-k,\uparrow(\downarrow)\rangle$, as shown in Fig.~\ref{pairing}(c,d). Thus, the Cooper pairs with parallel spins will oscillate spatially as $e^{-iqx}|\uparrow\uparrow\rangle+e^{iqx}|\downarrow\downarrow\rangle$, while the pairs  $|\uparrow\downarrow\rangle \pm |\uparrow\downarrow \rangle$ remain unchanged. Here we emphasize that the spin quantization axis aligns along different directions for different momenta, determined by the form of SOC in Fig.~\ref{pairing}(c). The spatially oscillatory pairs will decay after taking into account all possible wave vectors of $q$ \cite{Buzdin2005a} in the case of Fig.~\ref{pairing}(b).
Similarly, the triplet pairs $|\uparrow\uparrow\rangle$ and $|\downarrow\downarrow\rangle$ in Fig.~\ref{pairing}(c) will also generally decay rapidly in the SOC region. Therefore, in the presence of magnetization and generic SOC, only the pairs with zero center-of-mass momenta exhibit long-range proximity effect. However, there is an exception for a system with equal strengths of Rashba and Dresselhaus SOCs. In this case, the Fermi surfaces for two spin bands shifted in opposite directions by $Q=4m\alpha$, shown in Fig. \ref{pairing}(d). Here $m$ being the electron effective mass and $\alpha$ being the Rashba SOC strength. Thus, all of spatially oscillatory pairs have the same wave vector $Q$ and will not decay even in the presence of spin independent scattering. We show below that these oscillatory triplet pairs result in a long-range helical mode, dubbed "long-range spin-triplet helix'', in analogy to the persistent spin helix observed in two dimensional electron gases (2DEGs)\cite{Bernevig:2006_a,Stanescu:2007_a,LiuXin:2012_a,Weber:2007_a,Koralek:2009_a}. 

\begin{figure}
\includegraphics[width=0.8\columnwidth]{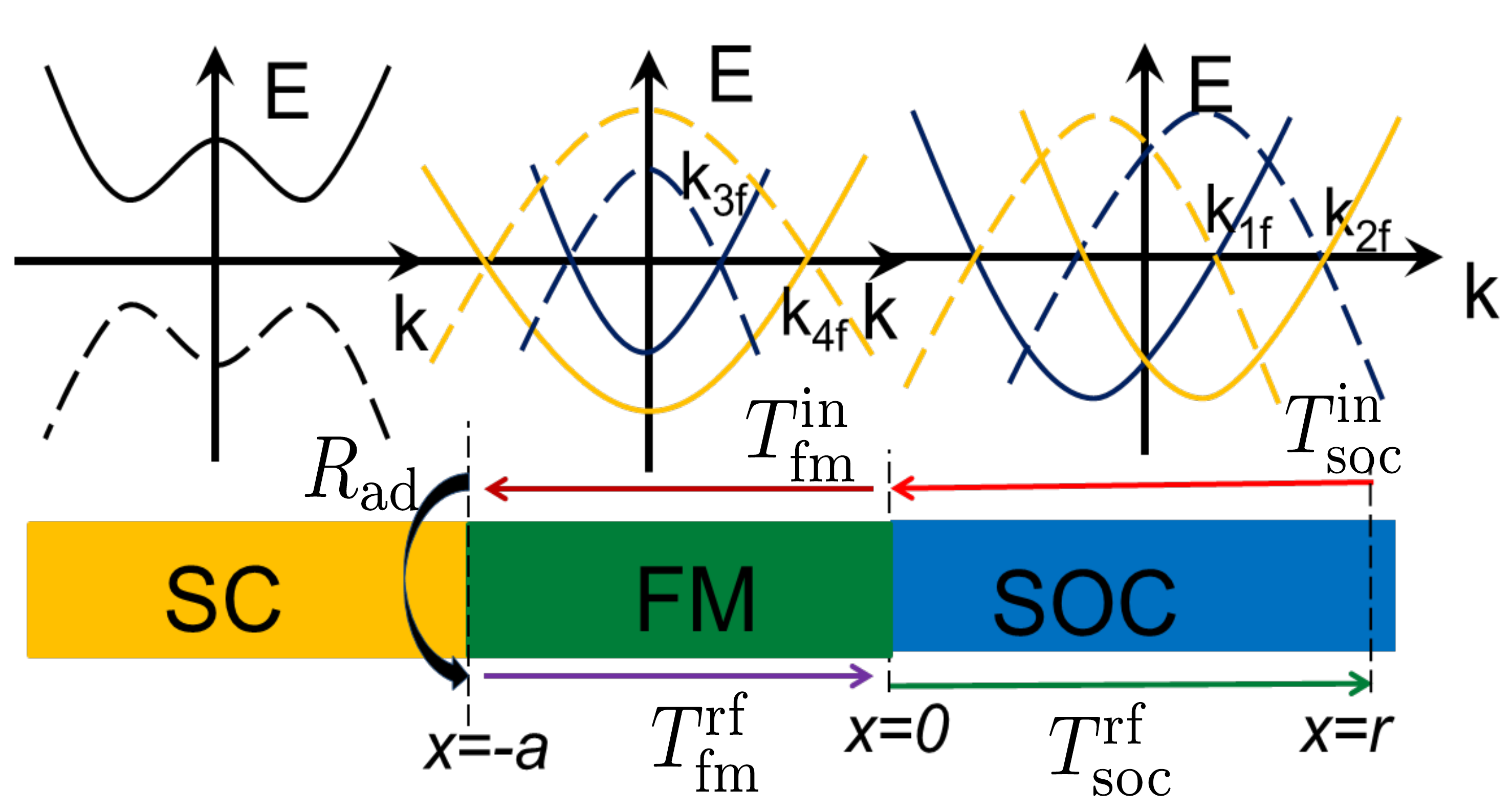}
\caption{ A schematic plot of a SC/FM/SOC junction. Energy dispersions for different regions are shown above the junction structure. The colors in the dispersion relation represent different spin indices and the solid lines (dashed lines) denote electron (hole) bands. $k_{1(2),f}$ and $k_{3(4),f}$ are the Fermi momenta of different spin bands for SOC and FM regions, respectively. Different propagation or reflection processes are denoted by $T^{\rm in(out)}_{\rm fm(soc)}$ or $R_{ad}$.}
\label{SFSOC}
\end{figure}

\begin{figure}
\centering
\begin{tabular}{l}
\includegraphics[width=0.8\columnwidth]{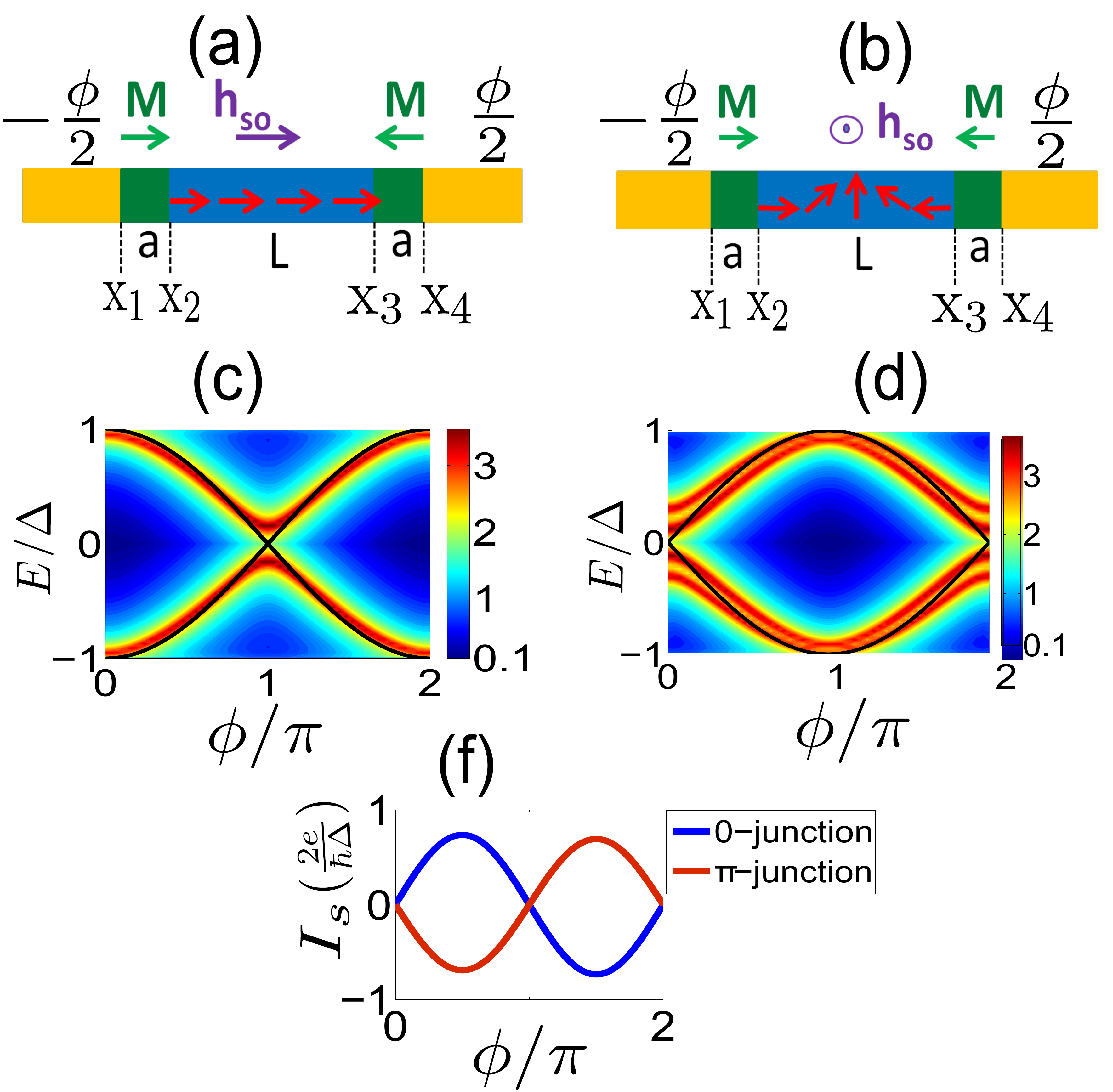}
\end{tabular}
\caption{The magnetization direction (the green arrows) and the effective magnetic field direction of SOC (the purple arrow) are shown (a) for a 0-junction and (b) a $\pi$-junction. The red arrows reveals the spatial distribution of d-vector. The phases of SCs at two sides are taken to be $\phi/2$ and $-\phi/2$. The color in (c) and (d) shows the spectral function of the SC/FM/SOC/FM/SC junction (logarithmic plot) as a junction of the relative phase $\phi$ for a 0-junction and $\pi$-junction, respectively. The black lines are the Andreev levels from analytical calculations. (f) shows the current-phase relation for the 0- and $\pi$-junction.}
\label{model}
\end{figure}

{\it Hamiltonian and pairing functions -}
We study the SC/normal-conductor structure whose Hamiltonian takes the form 
\begin{eqnarray*}
\hat{H}&=&\left(\begin{array}{cc}H_0& \hat{\Delta} \\ \hat{\Delta}^{\dagger} & -H_0^* \end{array}\right), \hat{\Delta}=\Delta(x) i\sigma_y, \nonumber \\ H_0&=&\left(\frac{\textbf p^2}{2m}-\mu \right)\sigma_0+\left(\textbf M(x) + \textbf h(x,\textbf k)\right) \cdot \bm \sigma,
\end{eqnarray*}
in the basis $[c_{\uparrow},c_{\downarrow},c^{\dag}_{\uparrow},c^{\dag}_{\downarrow}]^{\rm T}$, where $c_{\uparrow,\downarrow}$ and $c^{\dag}_{\uparrow,\downarrow}$ are electron annihilation and creation operators for different spins, $m$ is the electron mass, $\mu$ is the chemical potential, $\hat{\Delta}$ is the spin-singlet s-wave superconducting gap, $\textbf{M}$ is the magnetization, $\textbf{h}$ is the effective magnetic field of SOC and $\bm \sigma$ denotes the spin operators. The gap strength $\Delta(x)$ is zero in the proximity region and has a constant value $\Delta$ in the superconducting region. The magnetization $\textbf M(x)$ and effective magnetic field of SOC $\textbf h(x,k)$ are only present in the normal-conductor and depend on the spatial coordinate $x$ shown in Fig.~\ref{SFSOC} and Fig.~\ref{model}(a,b).

Cooper pairs in spin space can be described microscopically by a pairing function $f^R(E,\bm r)=(d_0\sigma_0+\textbf{d}\cdot \bm{\sigma})i\sigma_y$ \cite{Bergeret2001a,Champel2008}, which is the off diagonal block of the retarded Green's function \begin{eqnarray} \label{Pair-2}\left.	G^{\rm R}(E,\textbf r, \textbf r')\right|_{\textbf r= \textbf r'}=\left(\begin{array}{cc}g^{\rm R}(E,\textbf r)&f^{\rm R}(E,\textbf r) \\ \overline{f}^{\rm R}(E,\textbf r)&\overline{g}^{\rm R}(E,\textbf r) \end{array}\right). \end{eqnarray}  Here  $d_0$ and $\textbf{d}$ are the expectation value of singlet and triplet pairs respectively, $E$ is the energy, $\textbf{r}$ and $\textbf{r'}$ are the spatial coordinates; we have $f_{ij}^{\rm R}(E,\textbf r)=-(\overline{f}_{ij}^{\rm R}(-E,\textbf r))^{\dagger}$; and $g^{\rm R}$($\overline{g}^{\rm R}$) is the electron (hole) Green's function. Both $f^R$ and $g^R$ are $2 \times 2$ matrices in spin space. The superconducting gap is related to the pairing function by the equality $\hat{\Delta}=(1/2\pi)\int  dE \lambda f_E {\rm Im} f^{\rm R}$ where $\lambda$ is the attractive interaction strength and $f_E$ is the Fermi distribution. In the proximity region, the superconducting gap is zero because of $\lambda=0$, but the pairing function $f^{\rm R}$ can be nonzero. Below, we will calculate, in the presence of either magnetization or SOC, the spatial evolution of the pairing function $f^R(E,\textbf r)$ in the proximity region and show its consistence to the physical picture in Fig.~\ref{pairing}.

{\it $\textbf d$ vector in a one-dimensional (1D) SC/FM/SOC junction -}
In the ferromagnetic region ($x \in (-a,0)$) the SOC is zero, while in the SOC region ($x>0$) the magnetization is zero shown in Fig.~\ref{SFSOC}. In the SOC (FM) region, the Fermi wave vectors of the spin split bands, $k_{1\rm f}, k_{2\rm f}$ ($k_{3\rm f}, k_{4\rm f}$) in Fig~\ref{SFSOC}, satisfy
\begin{eqnarray}\label{split-1}
k_{2\rm f}-k_{1\rm f}=\frac{2|\textbf{h}(k_{\rm f})|}{\hbar v_{\rm f}}, \ \ k_{4\rm f}-k_{3\rm f}=\frac{2M}{\hbar v_{\rm f}},
 \end{eqnarray}
 with $\hbar v_{\rm f}=\hbar k_{\rm f}/m=\sqrt{2\mu/m}$, assuming $h,M\ll \mu$.
 The Green's functions $G^R$ can be related to the reflection matrix $R$ by the Fisher-Lee relation\cite{Fisher1981a} which has been applied to the superconducting proximity effect \cite{Lambert1993,Wang2001}. For 1D case, Fisher-Lee relation in the basis $[c_{\uparrow},c_{\downarrow},c^{\dag}_{\uparrow},c^{\dag}_{\downarrow}]^{\rm T}$ takes the form \cite{Lambert1993,Wang2001,Datta1997}
 \begin{eqnarray}\label{F-L-1}
 R_{ij}(E,\textbf r)=-\delta_{ij}+i\hbar\sqrt{v_i v_j}G_{ij}^R(E,\textbf r),
 \end{eqnarray}
 where $i,j=1,\dots,4$ and $v_{i(j)}$ is the velocity of the particle at energy $E$ in $i(j)$ channels. Therefore, we will calculate the reflection matrix to extract pairing functions in 1D case. For simplicity, we consider the clean limit with perfect transmission at FM/SOC boundary and ideal Andreev reflection at the FM/SC boundary.  The reflection matrix $R(x)$ in the SOC region can be decomposed into five matrices representing five steps shown in Fig.~\ref{SFSOC}: an electron first propagates from $x=r$ to the interface at $x=0$ ($T^{\rm in}_{\rm soc}$); it then propagates to the interface at $x=-a$ ($T^{\rm in}_{\rm fm}$); ideal Andreev reflection occurs at the SC/FM interface of $x=-a$ ($R_{\rm ad}$), where the electron is completely reflected as a hole; the reflected hole transmits back to $x=0$ ($T^{\rm rf}_{\rm fm}$), and finally to the SOC region at $x=r$ ($T^{\rm rf}_{\rm soc}$) \cite{SM_T}. Consequently, the scattering matrix $R(r)$ takes the form
\begin{eqnarray}\label{S-1}
R(r)=T^{\rm rf}_{\rm soc} T^{\rm rf}_{\rm fm}R_{\rm ad}T^{\rm in}_{\rm fm}T^{\rm in}_{\rm soc}.
\end{eqnarray}

When there is no SOC (i.e., $r=0$), the reflection matrix at FM/SOC boundary takes the form \cite{SM_T}
\begin{equation}\label{app:Andreev-2-0}R(r=0)=T^{\rm rf}_{\rm fm}R_{\rm ad}T^{\rm in}_{\rm fm}=-\hbar v_{\rm f} (d_0\sigma_0+\textbf{d} \cdot \bm{\sigma} )i\sigma_y\otimes \tau_y,
 \end{equation}
 where 
 \begin{equation}\label{dR-1}
 (d_0,\textbf{d})=-i\frac{e^{-i\alpha}}{\hbar v_{\rm f}}\left(\cos\left(\frac{2Ma}{\hbar v_{\rm f}}\right),i\sin\left(\frac{2Ma}{\hbar v_{\rm f}}\right) {\textbf m}\right),
\end{equation}
${\textbf m}={\textbf M}/M$, $\alpha=\arccos(E/\Delta)$ and $\tau_z=+1(-1)$ for the electron (hole) in the Nambu space. In the limit $M\ll \mu$, we take $\sqrt{v_i v_j}\approx v_{\rm f}$. Eqs. (\ref{app:Andreev-2-0}) and (\ref{dR-1}) show oscillation between singlet and triplet pairs as a function of $a$, the distance from the SC/FM interface. Thus, by choosing an appropriate length $a$ of the FM region, one can use the SC/FM junction to inject singlet or triplet pairs into the SOC region. 

When there is no FM ($a=0$), the reflection matrix reduces to $R(r)=T^{\rm rf}_{\rm soc} R_{\rm ad}T^{\rm in}_{\rm soc}=R_{ad}$ \cite{SM_T}  in the SOC region. This is because SOC does not lift the degeneracy of time reversed pairs, as shown in Fig~\ref{pairing} (c) and (d).
For an FM of length $a$ satisfying $2Ma/\hbar v_{\rm f}=\pi/2$, only triplet pairs with \textbf{d} vector along ${\textbf M}$ are injected into the SOC region. When the effective magnetic field of SOC is parallel to the magnetization, say $\textbf h(\textbf k) \parallel \textbf{M}$, the reflection matrix in the SOC region can be written as
\begin{eqnarray}\label{d-uniform}
R(r)=-e^{-i\alpha} {\textbf m}\cdot {\bm \sigma}i\sigma_y \otimes \tau_y.
 \end{eqnarray}
When $\textbf h(\textbf k) \perp \textbf M$, the reflection matrix in the SOC regime has the form
\begin{eqnarray}\label{d-helix}
R(r)=-\hbar v_{\rm f}(d_{1}\textbf m\cdot \bm{\sigma} +d_{2}\textbf m \times \textbf n \cdot \bm{\sigma}) i\sigma_y\otimes \tau_y, 
\end{eqnarray}
where
\begin{eqnarray}\label{d-helix-1}
\left(d_{1},d_{2}\right)=\frac{e^{-i\alpha}}{\hbar v_{\rm f}}\left(\cos(k_{2\rm f}-k_{1\rm f})r,\sin(k_{2\rm f}-k_{1\rm f})r\right), 
\end{eqnarray}
and ${\textbf n}$ is the unit direction of $\textbf{h}(k_{\rm f})$. Here $d_1$ and $d_2$ give the decomposition of the \vec{d}-vector along the direction ${\textbf m}$ and ${\textbf m\times \textbf n}$, respectively. 
Eq.~(\ref{d-uniform}) implies that \textbf{d} vector keeps its original direction in the case of $\textbf d \parallel \textbf h(\textbf k)$. In contrast, Eq.~(\ref{d-helix}) shows that in the case of $\textbf{d}\perp{\textbf h(\textbf k)}$, \textbf{d} vector precesses in the plane perpendicular to $\textbf h(\textbf k)$ when propagating along 1D SOC region. The above conclusions are consistent with our physical picture shown in Fig~\ref{pairing}(c,d).  Especially, based on Eq~(\ref{d-helix-1}), the precession of \textbf{d} vector leads to a helical structure, which is dubbed \textbf{d} helix or spin-triplet helix and schematically shown by red arrows in the SOC region of Fig~\ref{model} (b).

{\it  0 and $\pi$ Josephson junction transition -} 
To confirm the predicted \textbf{d} helix, we propose an experimental setup of a SC/FM/SOC/FM/SC junction (Fig. \ref{model}(a,b)) and show that the \textbf{d} helix can lead to a $0-\pi$ transition in Josephson junctions \cite{Buzdin2005a}. The magnetizations of two ferromagnetic layers point along $x$ and $-x$ direction (Fig.~\ref{model}(a,b)), to ensure a trivial 0-Josephson junction in the absence of the SOC region. The lengths of FMs are chosen to satisfy $2Ma/\hbar v_{\rm f}=\pi/2$, so only triplet pairs with $\textbf d$ vector along x direction are injected into the SOC region.

We consider two cases with the SOC $\textbf{h}(k)=\alpha k_x \hat{e}_x \parallel \textbf M$ in Fig.~\ref{model}(a) and $\textbf{h}(k)=\alpha k_x \hat{e}_y \perp \textbf M$ in Fig.~\ref{model}(b). The length of the SOC wire satisfy $(k_{2\rm f}-k_{1\rm f}) L=\pi$. To study the current-phase relation in this setup, we first calculate the Andreev levels numerically by evaluating the spectral function, $ {\rm Tr}[\sum_n g^R(E,x_n)]/N$, in a tight-binding model.  Here $g^{R}$ is the electron retarded Green's function defined in Eq.~(\ref{Pair-2}), $x_n$ represents the $n$th site and $N$ is the total number of sites in the proximity region. The spectral functions are plot as a function of the relative phase $\phi$ between two SCs in Fig.~3(c,d). The peaks shown by the red color indicate Andreev levels. We also obtain Andreev levels analytically using the standard scattering matrix method \cite{Schapers2001,Beenakker1992,SM_T}. The analytical results are shown by two black lines in Fig 3(c,d), which are consistent with the numerical results. It is noted that the crossings of the black curves at $\phi=\pi$, in Fig.~\ref{model}(c) and at $\phi= 0$; $2\pi$ in Fig.~\ref{model}(d) turn into anti-crossings in numerical results. This is because we impose a barrier potential at the SC/FM interfaces and include the Fermi velocity mismatch among different regions in numerical calculations, which remove all degeneracies in analytical results. The anti-crossing changes the period of the Josephson current at zero temperature, $I_s=\frac{2e}{\hbar}\sum_n \partial E_n/\partial \phi$ with the summation of negative Andreev levels, from $4\pi$ (black curves) to $2\pi$ \cite{Tang1997,Schapers2001}. The Josephson current for the Fig.~\ref{model}(c) gives the form of $I_s\sim \sin(\phi)$ (the blue line in Fig.~\ref{model}(f)), which corresponds to a 0-junction. In contrast, for the Fig. \ref{model}(d) we have $I_s\sim \sin(\phi+\pi)$ (the red line in Fig. \ref{model}(f)), indicating a $\pi$-junction. This 0-pi junction transition is consistent with the physical picture of the d-vector precession, shown by red arrows in Fig.~\ref{model}(a,b). Further calculations show that the $\pi$ junction is obtained for $L$ satisfying $\pi/2 <  (k_{1f}-k_{2f})L <3\pi/2$ \cite{Notpub}.

{\it  $\textbf d$ helix in a 2D system -} 
Having clarified the physics in a 1D model, we next ask if \textbf{d} helix also exists in a 2D system. For a 2DEG, the SOC has the form (assuming $x$-axis along $[110]$ direction)
\begin{eqnarray*}\label{SOC-2}
H_{\rm so}=(\alpha+\beta)k_x\sigma_y+(\beta-\alpha)k_y\sigma_x,
\end{eqnarray*}
where $\alpha$ and $\beta$ are the Rashba and Dresselhaus SOC strengths. 
When $\alpha=\beta$, the Fermi surface with the spin parallel (anti-parallel) to the y axis is shifted along $-x$ ($x$) direction by $Q/2$, as shown in Fig~\ref{pairing}(d). As a result, the eigenenergies of two spin states satisfy $\epsilon_{\rm 1}({\bf k})=\epsilon_{\rm 2}({\bf k+Q})$, where ${\bf Q}=4m\beta \hat{e}_x$, and $1$ $(2)$ denotes the spin parallel (anti-parallel) to the y axis. 
As shown in Refs.~\cite{Bernevig:2006_a,Weber:2007_a,Koralek:2009_a,Stanescu:2007_a,LiuXin:2012_a}, one can construct {\em spin} helix operators, which 
commute with the Hamiltonian and lead to a persistent spin helix mode. 

In our model with superconductivity, we can define triplet pairing operators
\begin{eqnarray}\label{dhelix-1}
	\hat{d}_{x}&=&\frac{1}{2}\left(\sum_{ \{ {\bf k},i\}} \delta(\epsilon_{ {\bf k},i}-\mu) c^{\dagger}_{ {\bf k},i} c^{\dagger}_{ {-\bf k}-(-1)^i {\bf Q},i}+\rm{h.c.}\right), \\ \label{dhelix-2}
	\hat{d}_{z}&=&\frac{1}{2i}\left(\sum_{ \{ {\bf k},i\}} \delta(\epsilon_{ {\bf k},i}-\mu) c^{\dagger}_{ {\bf k},i} c^{\dagger}_{ {-\bf k}-(-1)^i {\bf Q},i}-\rm{h.c.}\right)
 \end{eqnarray} 
where the summation is performed in the interval $\{\textbf{k},i\}=\{k_x<(-1)^i Q/2,k_y\}$ at the Fermi surface to avoid double counting. These two operators represent a \textbf{d} helix of triplet pairs with center-of-mass $\bf Q$ in x-z plane.
Since the operators $\hat{d}_{x,z}$ commute with the Hamiltonian $H_0+H_{so}$ \cite{SM_T}, a persistent \textbf{d} helix also exists in the triplet superconducting proximity region. It is also noted that in the case of $\alpha=\beta$, the Hamiltonian even with a spin independent scattering potential, $H=H_0+H_{\rm so}+V(\textbf r)\sigma_0$,  can be transformed to a Hamiltonian without SOC through the unitary matrix $U=\exp(-iQx/2)\sigma_y$. This is because $U$ is independent of momenta and commutes with $V(\textbf r)\sigma_0$. At the same time,  the triplet pairs with center-of-mass momentum $Q$ as defined in Eq. (\ref{dhelix-1}, \ref{dhelix-2}) is transformed to those with zero center-of-mass momentum as shown in Fig~\ref{pairing}(a). Therefore, we expect that this spin-triplet helix is immune to any spin-independent scattering and its decay length should be as long as the Cooper pairs coherence length \cite{Schapers2001} in the normal region. This can be further confirmed by solving Usadel equations \cite{Usadel:1970,Rammer:2007_a} with SOCs \cite{SM_T}.

\begin{figure}
\centering
\begin{tabular}{l}
\includegraphics[width=0.7\columnwidth]{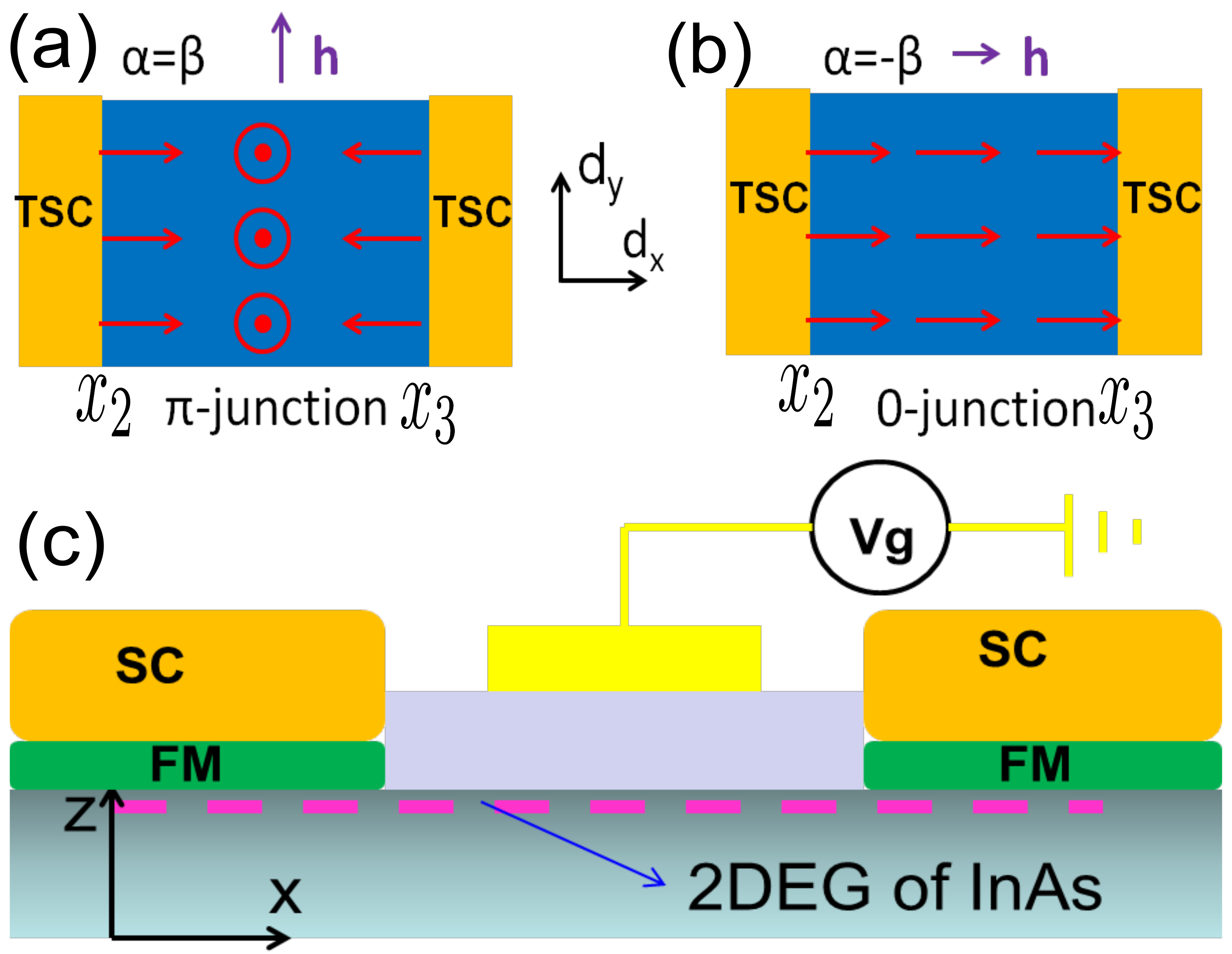}
\end{tabular}
\caption{The spatial dependence of ${\textbf d}$ vector of triplet pairs (the red arrows) and the corresponding effective magnetic field (the purple arrows) of the SOC are shown for (a) $\alpha=\beta$ ($\pi$-junction) and (b) $\alpha=-\beta$ (0-junction). TSC means triplet superconductor. (c)  The proposed 2D SC/FM/SOC/FM/SC structure for an electronic-tunable Josephson junction.}
\label{shift-trans}
\end{figure}

 In experiments, the Dresselhaus parameter $\beta$ is fixed while Rashba parameter $\alpha$ can be tuned by a gate voltage. Therefore, the following geometry can be used to confirm the oscillatory triplet pairs by observing an electrically tunable $0$-$\pi$ transition. 
The length $L$ of the SOC region is chosen to satisfy the condition $QL=\pi$. From the above discussion, when $\alpha=\beta$, the \textbf{d} vector of triplet pairs changes its sign after propagating from $x=x_2$ to $x=x_3$ (Fig.~\ref{shift-trans}(a)), leading to a $\pi$-junction. If we tune the Rashba parameter to $\alpha=-\beta$, the effective magnetic field of SOC $\textbf{h}=2\beta k_y \hat{e}_x$ is along the x direction, parallel to $\textbf{d}$ vector. Based on our theory, $\textbf d$ vector keeps its direction in the SOC region (Fig.~\ref{shift-trans}(b)) and we will have a $0$-junction. The proximity effect in the 2D Josephson junction for these two cases should be long-range according to our arguments. For realistic experiments, InAs quantum wells provide a potential candidate (Fig~\ref{shift-trans}(c)), because they show strong proximity effect due to their low Schottky barrier \cite{Doh2005a}. If the two FM layer are Ni, 1 nm thickness \cite{Klose2012a} is enough to convert singlet pairs in SC to triplet pairs on FM/InAs interface. For the effective mass $m_{\rm eff}=0.04\rm{\rm m}_{\rm{e}}$ and typical $\alpha=0.2 \rm{eV}$\AA in InAs quantum wells, we find $Q\approx  40\mu m^{-1}$, which corresponds to the length of $\sim80 nm$ of the SOC region to realize the Josephson $0-\pi$ junction transition. This length is much smaller than the coherence length, $\xi_{\rm N}=\hbar^2 \sqrt{2\pi n}/m_{\rm eff}2\pi k_{\rm B}T_c \approx 4 \mu\rm{m}$ \cite{Schapers2001}, where $n=10^{12} \rm{cm}^{-2}$ is the typical electron density in the InAs quantum well and $T_c=1.2K$ is the critical temperature of Al.

We acknowledge Yinghai Wu, Jimmy A. Hutasoit and Shou-Cheng Zhang for very helpful discussion. X.L. acknowledges partial support by the DOE under Grant No. DE-SC0005042.

%

\newpage

\begin{widetext}
\section{Supplementary material}

\setcounter{figure}{0}
\setcounter{equation}{0}
\renewcommand\thefigure{S\arabic{figure}}
\renewcommand\thetable{S\arabic{table}}
\renewcommand\theequation{S\arabic{equation}}

In the Supplementary Material, we provide details for the calculation of the propagation matrix and the Andreev levels for various geometries mentioned in the main text in the 1D clean limit. We also present the details for the spatial evolution of triplet pairs in the 2D system with general spin-orbit couplings (SOCs). Section I will derive the previously known results for superconductor/ferromagnet (SC/FM) junction \cite{SBuzdin2005a,SBergeret2005a} from the scattering matrix method, and Section II will consider SC/FM/SOC geometry. Section III will show how to calculate the Andreev levels in SC/FM/SOC/FM/SC junctions based on the scattering matrix method. The definition of the triplet pairing operators given in  Eqs. (10, 11) of the main text and their properties are given in Section IV. Section V describes the spatial evolution of the triplet pairs based on the Usadel equation.

\subsection{1D SC/FM junction}

We first consider how magnetization mixes different pairing functions in a one-dimensional (1D) ferromagnetic region of a SC/FM junction, schematically shown in Fig~\ref{S-M}.a. The effective Hamiltonian for this junction is given by
\begin{eqnarray}\label{Ham-5}
H_{\rm SC/FM}= \left(\begin{array}{cc}\left(\frac{\hat{P}^2}{2m}-\mu\right)\sigma_0&0\\0& -\left(\frac{\hat{P}^2}{2m}-\mu\right)\sigma_0\end{array}\right)+\Theta(x)\left(\begin{array}{cc}\bm{M}\cdot \bm \sigma&0\\0& -\bm{M}\cdot \bm \sigma^*\end{array}\right)+\Theta(-x) \left(\begin{array}{cc}0&\Delta i\sigma_y\\-\Delta i\sigma_y& 0\end{array}\right),
\end{eqnarray}
where $\Theta$ is the Heaviside step function, ${\bf M}$ denotes magnetization of FM, and $\Delta$ is the superconducting gap in the SC region. 
The SC/FM interface reflects incoming electrons (holes) into outgoing holes (electrons) and thereby induces a non-zero pairing function in the ferromagnetic region. To explore the spatial evolution of pairing function in a clean ferromagnetic wire, we formulate the reflection process by a matrix $R_{\rm fm}(a)$, given by
\begin{equation}
R_{\rm fm}(a)=T_{\rm fm}^{\rm rf}R_{\rm ad}T_{\rm fm}^{\rm{in}}
\end{equation}
 which is decomposed into three steps shown in Fig.~\ref{S-M}.a. An incoming electron (hole) is transmitted from $x=a$ to the SC/FM interface at $x=0$ ($T^{\rm in}_{\rm fm}$); then an ideal Andreev reflection occurs at SC/FM interface where the incoming electron (hole) is completely reflected to the outgoing hole (electron) ($R_{\rm ad}$); the reflected outgoing hole (electron) propagates back to $x=a$ ($T^{\rm rf}_{\rm fm}$).  We now calculate each factor separately.

\begin{figure}
\centering
\begin{tabular}{l}
\includegraphics[width=0.5\columnwidth]{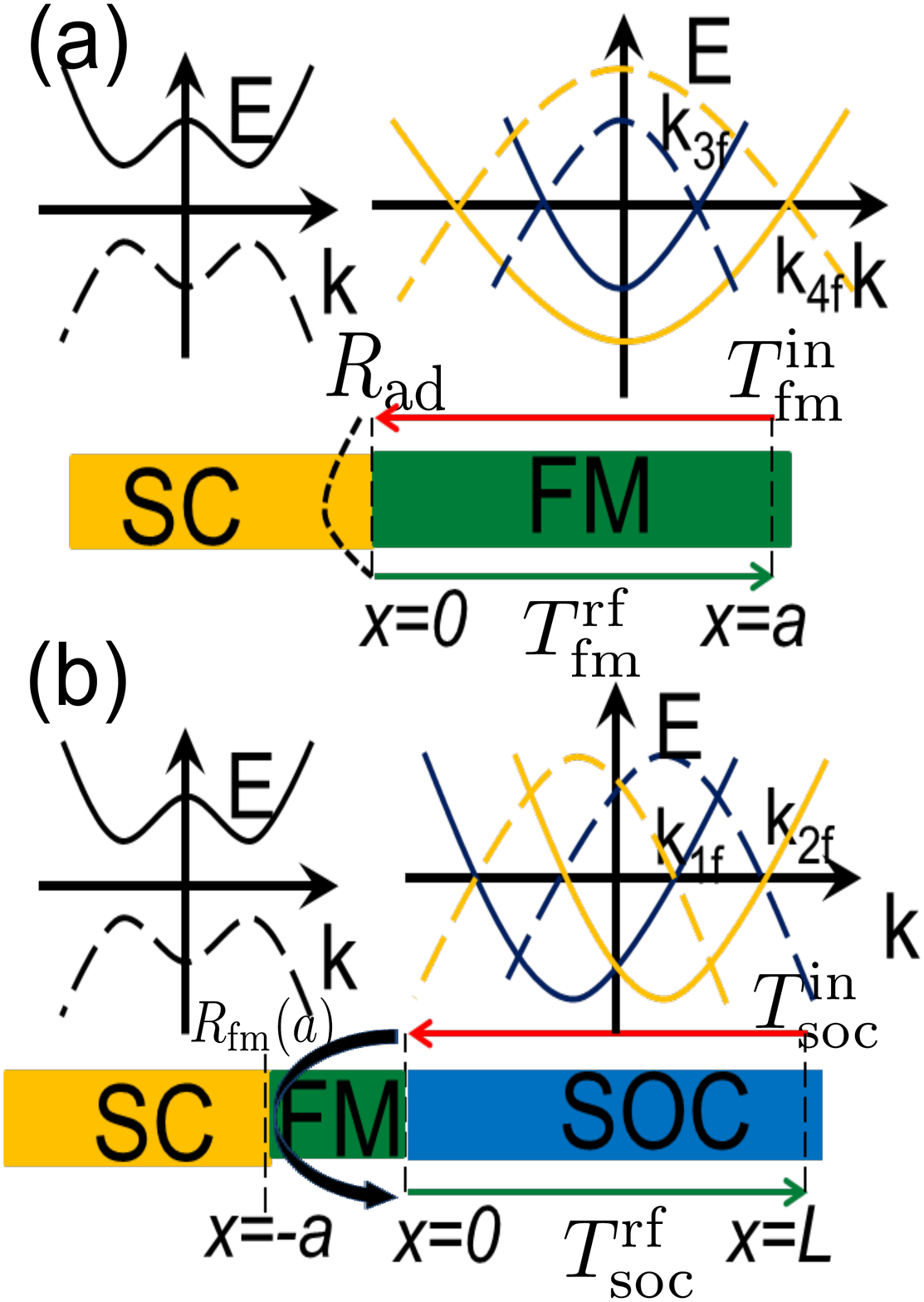}
\end{tabular}
\caption{Panels (a) and (b) show SC/FM and SC/FM/SOC junctions, respectively. The dispersions relations in various regions are shown; 
the colors represent the different spin indices and the solid lines (dashed lines) denote the electron (hole).  Different propagation processes defined as $T^{\rm in}_{\rm fm}$, $R_{\rm ad}$, $T^{\rm rf}_{\rm fm}$, $T^{\rm in}_{\rm soc}$ and $T^{\rm rf}_{\rm soc}$ are shown on the figure.  The quantities $k_{\rm 1 f}$, $k_{\rm 2f}$, $k_{\rm 3 f}$ and $k_{\rm 4f}$ are the Fermi momenta for different spin bands.}.
\label{S-M}
\end{figure}

In the first step, the wave functions of two incoming electrons and holes with opposite spins take the form 
\begin{eqnarray}\label{wf-1}
\Psi_{+\rm e}^{\rm in}=\left(\begin{array}{c} \psi_{+}\\ \hat{0} \end{array}\right)e^{- ik_{4 \rm f} x}, \ \ \ \Psi_{+h}^{\rm in}=\left(\begin{array}{c} \hat{0}\\ \psi_{+}^* \end{array}\right)e^{+ ik_{4 \rm f} x},\nonumber \\
\Psi_{- \rm e}^{\rm in}=\left(\begin{array}{c} \psi_{-}\\ \hat{0} \end{array}\right)e^{- ik_{3 \rm f} x}, \ \ \ \Psi_{- \rm h}^{in}=\left(\begin{array}{c} \hat{0}\\ \psi_{-}^* \end{array}\right)e^{+ ik_{3 \rm f} x},
\end{eqnarray}
where $\hat{0}=(0,0)^{\rm T}$, $k_{3\rm f}$ and $k_{4\rm f}$ (in Fig.~\ref{S-M}(a)) are the Fermi wave vectors of the minority and majority spin bands, respectively, and $\bm{M}\cdot \bm{\sigma}\psi_{\pm}=\pm |\bm M|\psi_{\pm}$. 
In the clean limit, the transmission matrix $T_{\rm fm}^{\rm in}$ describes the propagation of an incoming electron or hole from $x=a>0$ to the SC/FM interface and is given by  
\begin{eqnarray}\label{Tran-1}
T_{\rm fm}^{\rm{in}}&=&|\Psi_{+\rm e}^{\rm in}(0)\rangle \langle \Psi_{+\rm e}^{\rm in}(a)|+|\Psi_{+\rm h}^{\rm in}(0)\rangle \langle \Psi_{+\rm h}^{\rm in}(a)|+|\Psi_{-\rm e}^{\rm in}(0)\rangle \langle \Psi_{-\rm e}^{\rm in}(a)|+|\Psi_{-\rm h}^{\rm in}(0)\rangle \langle \Psi_{-\rm h}^{\rm in}(a)|\nonumber \\
&=&\frac{1}{2}\left(\begin{array}{cc} e^{ik_{4 \rm f}a}(\sigma_0+\bm m \cdot \bm\sigma)&0\\0&e^{-ik_{4 \rm f}a}(\sigma_0+\bm m \cdot\bm\sigma^*) \end{array}\right)+\frac{1}{2}\left(\begin{array}{cc} e^{ik_{3 \rm f}a}(\sigma_0-\bm m \cdot\bm\sigma)&0\\0&e^{-ik_{3 \rm f}a}(\sigma_0-\bm m \cdot\bm\sigma^*) \end{array}\right),\nonumber \\
&=&\left(\begin{array}{cc} e^{i\beta}U_i&0\\0& e^{-i\beta} U_i^* \end{array}\right).
\end{eqnarray}
Here $\beta=(k_{4\rm f}+k_{3 \rm f})a/2$ and
\begin{eqnarray}\label{Tran-1-1}
U_{\rm fm}&=&\exp(i\frac{(k_{4 \rm f}-k_{3 \rm f})a}{2}\bm m \cdot \bm \sigma)=\cos(\frac{(k_{4 \rm f}-k_{3 \rm f})}{2}a)\sigma_0+i\sin(\frac{(k_{4 \rm f}-k_{3 \rm f})}{2}a)\bm m \cdot \bm \sigma,\nonumber \\
U_{\rm fm}^{\dagger}&=&\exp(-i\frac{(k_{4 \rm f}-k_{3 \rm f})a}{2}\bm m \cdot \bm \sigma)=\cos(\frac{(k_{4 \rm f}-k_{3 \rm f})}{2}a)\sigma_0-i\sin(\frac{(k_{4 \rm f}-k_{3 \rm f})}{2}a)\bm m \cdot \bm \sigma,\nonumber \\
U_{\rm fm}^{*}&=&\exp(-i\frac{(k_{4 \rm f}-k_{3 \rm f})a}{2}\bm m \cdot \bm \sigma^{*})=\cos(\frac{(k_{4 \rm f}-k_{3 \rm f})}{2}a)\sigma_0-i\sin(\frac{(k_{4 \rm f}-k_{3 \rm f})}{2}a) \bm m \cdot \bm\sigma^{*},\nonumber \\
U_{\rm fm}^{\rm T}&=&\exp(i\frac{(k_{4 \rm f}-k_{3 \rm f})a}{2} \bm m \cdot \bm \sigma^{*})=\cos(\frac{(k_{4 \rm f}-k_{3 \rm f})}{2}a)\sigma_0+i\sin(\frac{(k_{4 \rm f}-k_{3 \rm f})}{2}a)\bm m \cdot \bm \sigma^*.\nonumber \\
\end{eqnarray}
In the second step, the ideal Andreev reflection matrix has the form \cite{SSchapers2001}
\begin{eqnarray}\label{Andreev-1}
R_{\rm ad}=e^{-i\alpha}\left(\begin{array}{cc}0&i\sigma_y\\ -i\sigma_y&0 \end{array}\right),
\end{eqnarray}
where $\alpha=\arccos(E/\Delta)$ with the energy $E$ satisfying $|E|<\Delta$. 
In the third step, there are four outgoing particles in the ferromagnetic region, with wave functions given by
 \begin{eqnarray}\label{wf-5}
\Psi_{+\rm e}^{\rm rf}=\left(\begin{array}{c} \psi_{+}\\ \hat{0} \end{array}\right)e^{ ik_{4 \rm f} x}, \ \ \ \Psi_{+h}^{\rm rf}=\left(\begin{array}{c} \hat{0}\\ \psi_{+}^* \end{array}\right)e^{- ik_{4\rm f} x},\nonumber \\
\Psi_{- \rm e}^{\rm rf}=\left(\begin{array}{c} \psi_{-}\\ \hat{0} \end{array}\right)e^{ ik_{3\rm f} x}, \ \ \ \Psi_{- \rm h}^{\rm rf}=\left(\begin{array}{c} \hat{0}\\ \psi_{-}^* \end{array}\right)e^{- ik_{3\rm f} x}.
 \end{eqnarray}
The transmission matrix $T^{\rm rf}_{\rm fm}$ describes the outgoing waves moving back from the SC/FM interface at $x=0$ to $x=a$, given by 
 \begin{eqnarray}\label{Tran-1-2}
T_{\rm fm}^{\rm rf}&=&|\Psi_{+\rm e}^{\rm rf}(a)\rangle \langle \Psi_{+\rm e}^{\rm rf}(0)|+|\Psi_{+\rm h}^{\rm rf}(a)\rangle \langle \Psi_{+\rm h}^{\rm rf}(0)|+|\Psi_{-\rm e}^{\rm rf}(a)\rangle \langle \Psi_{-\rm e}^{\rm rf}(0)|+|\Psi_{-\rm h}^{\rm rf}(a)\rangle \langle \Psi_{-\rm h}^{\rm rf}(0)|\nonumber \\&=&\frac{1}{2}\left(\begin{array}{cc} e^{ik_{4 \rm f}a}(\sigma_0+\bm m \cdot \bm \sigma)&0\\0&e^{-ik_{4 \rm f}a}(\sigma_0+\bm m \cdot \bm \sigma^*) \end{array}\right)+\frac{1}{2}\left(\begin{array}{cc} e^{ik_{3 \rm f}a}(\sigma_0-\bm m \cdot \bm\sigma)&0\\0&e^{-ik_{3 \rm f}a}(\sigma_0-\bm m \cdot \bm\sigma^*)\end{array}\right)\nonumber \\
&=&\left(\begin{array}{cc} e^{i\beta}U_i&0\\0& e^{-i\beta} U_i^* \end{array}\right),
\end{eqnarray}
It is noted that $T^{\rm in}_{\rm fm}$ has the same form to $T^{\rm rf}_{\rm fm}$, which is consistent to the fact that the magnetization respects the inversion symmetry. The total reflection matrix at $x=a$ in the ferromagnetic region is then given by 
\begin{eqnarray}\label{app:Andreev-2-0}R_{\rm fm}(a)&=&T_{\rm fm}^{\rm rf}R_{\rm ad}T_{\rm fm}^{\rm{in}}\nonumber \\
&=&\left(\begin{array}{cc} e^{i\beta}U_i&0\\0& e^{-i\beta} U_i^{*}\end{array}\right)e^{-i\alpha}\left(\begin{array}{cc}0&i\sigma_y\\ -i\sigma_y&0 \end{array}\right)\left(\begin{array}{cc} e^{i\beta}U_i&0\\0& e^{-i\beta} U_i^{*}\end{array}\right)\nonumber \\
&=&e^{-i\alpha}\left(\begin{array}{cc}0&U_ii\sigma_yU_i^{*}\\ -U_i^{*}i\sigma_y U_i&0 \end{array}\right)\nonumber \\
&=&\left(\begin{array}{cc}0&\cos((k_{4 \rm f}-k_{3 \rm f})a)i\sigma_y+i\sin((k_{4 \rm f}-k_{3 \rm f})a)\bm m\cdot \bm \sigma i\sigma_y\\ \cos((k_{4 \rm f}-k_{3 \rm f})a)(-i\sigma_y)+i\sin((k_{4 \rm f}-k_{3 \rm f})a)\bm m\cdot \bm \sigma^* i\sigma_y&0 \end{array}\right)\nonumber \\
&=&e^{-i\alpha}\left[\cos(k_{4 \rm f}-k_{3 \rm f})a \left(\begin{array}{cc}0&i\sigma_y\\-i\sigma_y&0 \end{array}\right)+i\sin(k_{4 \rm f}-k_{3 \rm f})a\left(\begin{array}{cc}0&\bm m \cdot \bm \sigma i\sigma_y\\(\bm m \cdot \bm \sigma i\sigma_y)^{\dagger}&0 \end{array}\right)
 \right]
 \end{eqnarray}
 The pairing function can now be obtained by the Fisher-Lee relation \cite{SFisher1981a} shown in the main text, and is given by 
 \begin{eqnarray}\label{dR-1}
 f^R(E,x)&=&(d_0\sigma_0+\bm{d}\cdot \bm{\sigma})i\sigma_y,\nonumber \\
 d_0&=&-i\frac{e^{-i\alpha}}{\hbar v_{\rm f}}\cos((k_{4 \rm f}-k_{3 \rm f})a),\ \  \bm d=\frac{e^{-i\alpha}}{\hbar v_{\rm f}}\sin((k_{4 \rm f}-k_{3 \rm f})a)\bm m.
\end{eqnarray}
The above equations display the spatial oscillation between singlet and triplet pairs in the FM region.  We note that the $\bm{d}$-vector is along the direction of magnetization $\bm{M}$.

\subsection{1D SC/FM/SOC junction}

We next consider a 1D SC/FM/SOC junction. The calculation is conceptually similar to that given above, although the details are more complicated. The reflection matrix is now given by
\begin{equation}
R_{\rm soc}(L)=T_{\rm soc}^{\rm rf}R_{\rm fm}(a)T_{\rm soc}^{\rm{in}}
\end{equation}
where $R_{\rm fm}(a)$ has already been calculated above.
The SC/FM junction is utilized as a source of singlet and triplet pairs, and the relative strengths of the two can be tuned by varying $a$, the length of the FM region.  

The Hamiltonian in the SOC wire has the form
\begin{eqnarray}\label{Ham-5-1}
H_{\rm SOC}&=& \left(\begin{array}{cc}\left(\frac{\hat{p}^2}{2m}-\mu\right)\sigma_0&0\\0& -\left(\frac{\hat{p}^2}{2m}-\mu\right)\sigma_0\end{array}\right)+\left(\begin{array}{cc}\bm{h(\hat{p})}\cdot \bm \sigma&0\\0& -\bm{h^*(\hat{p})}\cdot \bm \sigma^*\end{array}\right)
\end{eqnarray}
where $\bm{\hat{p}}$ is the momentum operator,
$m$ is the electron mass, $\bm{h}(\bm{\hat p})$ is the effective magnetic field due to SOC and $\mu$ is the
chemical potential. In the SOC region, the incoming electrons and holes propagate to the FM/SOC interface with the wave functions
\begin{eqnarray}\label{wf-1-1}
\Psi_{+e}^{\rm in}=\left(\begin{array}{c} \psi_{+}\\ \hat{0} \end{array}\right)e^{-ik_{1 \rm f} x}, \ \ \ \Psi_{+h}^{\rm in}=\left(\begin{array}{c} \hat{0}\\ \psi_{+}^* \end{array}\right)e^{ ik_{2 \rm f} x},\nonumber \\
\Psi_{-e}^{\rm in}=\left(\begin{array}{c} \psi_{-}\\ \hat{0} \end{array}\right)e^{-ik_{2 \rm f} x}, \ \ \ \Psi_{-h}^{\rm in}=\left(\begin{array}{c} \hat{0}\\ \psi_{-}^* \end{array}\right)e^{ ik_{1 \rm f} x},
\end{eqnarray}
where $k_{2\rm f}$ and $k_{1 \rm f}$ (in Fig~\ref{S-M}(b)) are the Fermi wave vectors of the majority and minority spin bands respectively and  $\bm n\cdot\bm{\sigma}\psi_{\pm}=\pm\psi_{\pm}$ with $\bm n={\bf h}/{|\bf h|}$. The wave functions of outgoing electrons and holes are given by 
 \begin{eqnarray}\label{wf-6-1}
\Psi_{+\rm e}^{\rm rf}=\left(\begin{array}{c} \psi_{+}\\ \hat{0} \end{array}\right)e^{ ik_{2 \rm f} x}, \ \ \ \Psi_{+h}^{\rm rf}=\left(\begin{array}{c} \hat{0}\\ \psi_{+}^* \end{array}\right)e^{- ik_{1 \rm f} x},\nonumber \\
\Psi_{- \rm e}^{\rm rf}=\left(\begin{array}{c} \psi_{-}\\ \hat{0} \end{array}\right)e^{ ik_{1 \rm f} x}, \ \ \ \Psi_{- \rm h}^{\rm rf}=\left(\begin{array}{c} \hat{0}\\ \psi_{-}^* \end{array}\right)e^{- ik_{2 \rm f} x},
 \end{eqnarray}
To obtain the pairing function in the SOC wire, we consider a perfect contact at FM/SOC interface. The transmission from $x=L$ to the SC/FM interface at $x=0$ is represented by the matrix
\begin{eqnarray}\label{Tran-1-5}
T_{\rm soc}^{\rm{in}}&=&\frac{1}{2}\left(\begin{array}{cc} e^{ik_{2 \rm f}L}(\sigma_0+\bm n \cdot \bm \sigma)&0\\0&e^{-ik_{2 \rm f}L}(\sigma_0+\bm n \cdot \bm \sigma^*) \end{array}\right)+\frac{1}{2}\left(\begin{array}{cc} e^{ik_{1 \rm f}L}(\sigma_0-\bm n \cdot \bm \sigma)&0\\0&e^{-ik_{1 \rm f}L}(\sigma_0-\bm n \cdot \bm \sigma^*) \end{array}\right)\nonumber \\
&=&\left(\begin{array}{cc} e^{i\beta}U_i&0\\0& e^{-i\beta} U_i^* \end{array}\right)
\end{eqnarray}
where $\beta=(k_{2 \rm f}+k_{1 \rm f})L/2$ and
\begin{eqnarray}\label{Tran-1-3}
U_{\rm soc}&=&\exp(i\frac{(k_{2 \rm f}-k_{1 \rm f})L}{2}\bm n \cdot \bm \sigma)=\cos(\frac{(k_{2 \rm f}-k_{1 \rm f})}{2}L)\sigma_0+i\sin(\frac{(k_{2 \rm f}-k_{1 \rm f})}{2}L)\bm n \cdot \bm \sigma,\nonumber \\
U^{\dagger}_{\rm soc}&=&\exp(-i\frac{(k_{2 \rm f}-k_{1 \rm f})L}{2}\bm n \cdot \bm \sigma)=\cos(\frac{(k_{2 \rm f}-k_{1 \rm f})}{2}L)\sigma_0-i\sin(\frac{(k_{2 \rm f}-k_{1 \rm f})}{2}L)\bm n \cdot \bm \sigma,\nonumber \\
U^{*}_{\rm soc}&=&\exp(-i\frac{(k_{2 \rm f}-k_{1 \rm f})L}{2}\bm n \cdot \bm \sigma^{*})=\cos(\frac{(k_{2 \rm f}-k_{1 \rm f})}{2}L)\sigma_0-i\sin(\frac{(k_{2 \rm f}-k_{1 \rm f})}{2}L)\bm n \cdot \bm \sigma^{*},\nonumber \\
U^{\rm T}_{\rm soc}&=&\exp(i\frac{(k_{2 \rm f}-k_{1 \rm f})L}{2}\bm n \cdot \bm \sigma^{*})=\cos(\frac{(k_{2 \rm f}-k_{1 \rm f})}{2}L)\sigma_0+i\sin(\frac{(k_{2 \rm f}-k_{1 \rm f})}{2}L)\bm n \cdot \bm \sigma^*.\nonumber \\
\end{eqnarray}
The reflected hole (electron) moves back from the interface to $x=L$, which is represented by the matrix
 \begin{eqnarray}\label{Tran-1-4}
T_{\rm soc}^{\rm rf}&=&\frac{1}{2}\left(\begin{array}{cc} e^{ik_{1 \rm f}L}(\sigma_0+\bm n \cdot \bm \sigma)&0\\0&e^{-ik_{1 \rm f}L}(\sigma_0+\bm n \cdot \bm \sigma^*) \end{array}\right)+\frac{1}{2}\left(\begin{array}{cc} e^{ik_{2 \rm f}L}(\sigma_0-\bm n \cdot \bm \sigma)&0\\0&e^{-ik_{2 \rm f}L}(\sigma_0-\bm n \cdot \bm \sigma^*)\end{array}\right)\nonumber \\
&=&\left(\begin{array}{cc} e^{i\beta}U_i^{\dagger}&0\\0& e^{-i\beta} U_i^T \end{array}\right). 
\end{eqnarray}
It is noted that $T_{\rm soc}^{\rm rf}$, describing the propagation away from the F/SOC interface, is different from $T^{\rm in}_{\rm soc}$, describing the propagation towards the F/SOC interface. This indicates the fact that SOC breaks inversion symmetry.

We now have all the information needed to evaluate $R_{\rm soc}(L)$, and hence the pairing function. We specialize below to the case  
$(k_{4\rm f}-k_{3\rm f})a=\pi/2$, for which, according to Eq.~\ref{app:Andreev-2-0}, the SC/FM junction behaves as a reservoir of only triplet pairs whose $d$-vector is along the magnetization direction. When the magnetization ${\bf M}$ in the ferromagnetic region is parallel to ${\bf h}$ in the SOC region, the reflection matrix at $x=L$ in the SOC region is the same as that in the ferromagnetic region
\begin{eqnarray}\label{Andreev-2}
R_{\rm soc}(L)&=&T_{\rm soc}^{\rm rf}R_{\rm fm}(a)T_{\rm soc}^{\rm{in}}=R_{\rm fm}(a),
\end{eqnarray}
which implies that the SOC will not affect the triplet pair whose \vec{d}-vector is parallel to the effective magnetic field of SOC.
When ${\bf M}$ is perpendicular to ${\bf h}$, the refection matrix shows an oscillating behavior \begin{eqnarray}\label{Andreev-2-1}
R_{\rm soc}(L)= ie^{-i\alpha}\left[\cos(k_{2 \rm f}-k_{1 \rm f})L\left(\begin{array}{cc}0&\bm m \cdot \bm \sigma i\sigma_y\\(\bm m \cdot \bm \sigma i\sigma_y)^{\dagger}&0 \end{array}\right)+\sin(k_{2 \rm f}-k_{1 \rm f})L \left(\begin{array}{cc}0&\bm m \times \bm n \cdot \bm \sigma i\sigma_y\\(\bm m \times \bm n \cdot \bm \sigma i\sigma_y)^{\dagger}&0 \end{array}\right)
 \right] \nonumber
\end{eqnarray}
which is identical to rotate the triplet pair in the plane perpendicular to $\bm{h}$. 

\subsection{Scattering matrix method in SC/FM/SOC/FM/SC junction}

We now show the scattering matrix method in the SC/FM/SOC/FM/SC junction.
The magnetizations of two ferromagnetic layers point along $x$ and $-x$ direction (Fig.3(a,b) in the main text), to ensure a trivial 0-Josephson junction in the absence of the SOC region. The lengths of FMs are chosen to satisfy $2Ma/\hbar v_{\rm f}=\pi/2$, so only triplet pairs with $\bm d$-vector along x direction are injected into the SOC region based on Eq.~(\ref{dR-1}). From Eq.~(\ref{app:Andreev-2-0},\ref{dR-1}), the associated reflection matrix at the interface of $x_2$ ($x_3$) takes the form
 \begin{eqnarray}\label{AR-1}
 R_{\rm fm}^{-(+)}=ie^{-i\alpha}\left(\begin{array}{cc}0& e^{-(+)i\phi/2}\sigma_x i\sigma_y\\(e^{+(-)i\phi}\sigma_x i\sigma_y)^{\dagger}&0 \end{array}\right).
 \end{eqnarray}
where $-\phi/2$ ($\phi/2$) is the phase of the left (right) superconductor. The discrete Andreev levels in the Josephson junction can be obtained from the condition \cite{SSchapers2001,SBeenakker1992}
\begin{eqnarray}\label{Andreev level-0}
\rm{Det}\left(I_{4\times 4}-T_{\rm soc}^{\rm in}R_{\rm fm}^{-}T_{\rm soc}^{\rm rf}R_{\rm fm}^{+}\right)=0,
\end{eqnarray}
where $I_{4\times 4}$ is a 4 by 4 identity matrix. When the effective magnetic field of SOC is along y direction, substituting Eq. (S15,S17) into Eq. (S20) and taking $k_{\rm 2f}-k_{\rm 1f}=\pi$, we have
\begin{eqnarray*}
{\rm Det} (I_{4\times 4}-T_{\rm soc}^{\rm in}R_{\rm fm}^{-}T_{\rm soc}^{\rm rf}R_{\rm fm}^{+})={\rm Det}\left(\begin{array}{cc} (1+e^{-2i\alpha-i\phi})\sigma_0 & 0 \\ 0 &(1+e^{-2i\alpha+i\phi})\sigma_0 \end{array}\right)=0,
\end{eqnarray*}
which gives the two-fold degenerate Andreev levels $E=\pm \Delta \cos(\frac{\phi+\pi}{2})$. When the effective magnetic field of SOC is along x or -x direction, we have 
\begin{eqnarray*}
{\rm Det} (I_{4\times 4}-T_{\rm soc}^{\rm in}R_{\rm fm}^{-}T_{\rm soc}^{\rm rf}R_{\rm fm}^{+})={\rm Det}\left(\begin{array}{cc} (1-e^{-2i\alpha-i\phi})\sigma_0 & 0 \\ 0 &(1-e^{-2i\alpha+i\phi})\sigma_0 \end{array}\right)=0,
\end{eqnarray*}
which gives $E=\pm \Delta \cos(\frac{\phi}{2})$.
\subsection{Persistent triplet helix}
 In the two dimensional case, SOC in general induces a destructive interference of different transverse modes shown in Fig 1(c) in the main text. This will lead to a rapid decay of triplet pairing function. However, for some particular forms of SOC, triplet pairs can precess in a coherent way, resulting in a long range proximity effect. Below, we will show how to achieve a long range proximity effect of triplet pairing functions in a 2DEG system. 2DEGs usually possess two kinds of SOCs, namely the Rashba and Dresselhaus terms, given by
\begin{eqnarray}\label{SOC-2}
	H_{\rm R\&D}=H_{\rm Rashba}+H_{\rm Dresselhaus}=\alpha(k_x\sigma_y-k_y\sigma_x)+\beta(k_x\sigma_x-k_y\sigma_y),
\end{eqnarray}
where $\alpha$ and $\beta$ are the coefficients of Rashba and Dresselhaus SOCs, respectively. In the case of $\alpha=\beta$, the SOC Hamiltonian takes the form 
\begin{eqnarray}\label{SOC-3}
H_{\rm R\&D}=\alpha(k_x-k_y)(\sigma_x+\sigma_y).
 \end{eqnarray} 
 The form of the Hamiltonian is simplified if we re-define $(k_x-k_y)/\sqrt{2}\rightarrow k_x$ and $(\sigma_x+\sigma_y)/\sqrt{2}\rightarrow \sigma_z$ to $H_{\rm R\&D}=2\alpha k_x \sigma_z$. The two spin bands with opposite spins are shifted in the opposite directions. The Hamiltonian in the spin and Nambu space has the form
 \begin{eqnarray}\label{Ham-3}
H=\frac{1}{2}\sum_{\bm k,i}\left(\begin{array}{c} c_{\bm k,i}^{\dagger} \\ c_{-k,i}  \end{array} \right)^{\rm T} \left( \begin{array}{cc} k^2/2m-(-1)^iQ k_x-\mu  & 0 \\ 0&-( k^2/2m-(-1)^iQ k_x-\mu) \end{array} \right) \left(\begin{array}{c} c_{\bm k,i} \\ c_{-k,i}^{\dagger} \end{array} \right)
 \end{eqnarray} 
 where $i=1 (2)$ is for the spin parallel (anti-parallel) to $z$ direction, we define $Q=4m\alpha$, and $m$ is the effective mass of 2DEGs. 
 We construct the triplet helix operators as
\begin{eqnarray}\label{SOC-3}
\hat{d}^{-}&=&\sum_{\{\bm k, i\}} \delta(\epsilon_{ {\bf k},i}-\mu)c_{ {-\bf k}-(-1)^i {\bf Q},i}c_{\bm{k,i}}, 
\\ \label{SOC-4} \hat{d}^{+}&=&\sum_{\{\bm k, i\}} \delta(\epsilon_{ {\bf k},i}-\mu)c^{\dagger}_{ {\bf k},i} c^{\dagger}_{ {-\bf k}-(-1)^i {\bf Q},i},
\end{eqnarray}
where
\begin{equation}
	\epsilon_{k,i}=\frac{k^2-(-1)^i Q k_x}{2m}
	\label{eq:relation1}
\end{equation}
and the summation is performed over the Fermi surface, where we choose the interval $\{\bm k,i\}=\{k_x<(-1)^i Q/2,k_y\}$ to avoid double counting.
Due to the dispersion relation in Eq.~\ref{eq:relation1}, $\hat{d}^{\pm}$ defined in Eq.~(\ref{SOC-3},\ref{SOC-4}) commute with the Hamiltonian in Eq.~\ref{Ham-3}  
\begin{eqnarray}\label{SOC-5}
[H,\hat{d}^{-}]=\sum_{\{\bm k, i\}} \delta(\epsilon_{\bm{k},i}-\mu)\left(\epsilon_{-\bm{k}-(-1)^i\bm{Q},i}-\epsilon_{\bm{k},i}\right)c_{-\bm{k}-(-1)^i\bm{Q},i}c_{\bm{k}}=0,
\end{eqnarray}
\begin{eqnarray}\label{SOC-6}
[H,\hat{d}^{+}]=\sum_{\{\bm k, i\}} \delta(\epsilon_{\bm{k},i}-\mu)\left(\epsilon_{-\bm{k}-(-1)^i\bm{Q},i}-\epsilon_{\bm{k},i}\right)c_{\bm{k}}^{\dagger}c^{\dagger}_{-\bm{k}-(-1)^i\bm{Q},i}=0. 
\end{eqnarray}
which is the reason why SOC with $\alpha=\beta$ does not cause a decay of the triplet $\vec{d}$ helix for the center-of-mass momentum $Q$.

   To further confirm the long range triplet order in the dirty limit, we derive the Usadel equation in the proximity region with isotropic spin-independent scattering time $\gamma$. First, we derive the dynamic equation of the annihilation operator
   \begin{equation}\label{anih-1}
    {\psi}_{\mu}(t,\bm x)=e^{i\hat{H}t}\hat{\psi}_{\mu}(\bm x)e^{-i\hat{H}t}=e^{i\hat{H} t}\left(\sum_{\bm k}\hat{\psi}_{\mu}(\bm k)e^{i\bm k \cdot \bm x}\right)e^{-i\hat{H}t}
   \end{equation}
   where $\hat{H}=\sum_{\mu,\nu,\bm k}\psi^{\dagger}_{\mu}(\bm k) H_{\mu \nu}(\bm k)\psi_{\nu}(\bm k)$ and
   \begin{eqnarray}\label{Usadel-2}
   H(\bm k)=\left(\begin{array}{cc} \frac{k^2}{2m}&0 \\ 0& \frac{k^2}{2m} \end{array}\right)+\bm M \cdot \bm \sigma+\bm h(\bm k)\cdot \bm \sigma,
   \end{eqnarray}
   where $\bm M$ is the magnetization and $\bm h(\bm k)$ is the SOC field. Therefore we have
   \begin{eqnarray}\label{Dy-1}
   i\partial_t \hat{\psi}_{\mu}(t,\bm x)&=&e^{i\hat{H}t}\left[\sum_{\bm k}\hat{\psi}_{\mu}(\bm k)e^{i\bm k \cdot \bm x},\hat{H} \right] e^{-i\hat{H}t}\nonumber \\
   &=& e^{i\hat{H}t}\sum_{\bm{k,k'},\lambda,\nu}\left[\hat{\psi}_{\mu}(\bm k)e^{i\bm k \cdot \bm x},\psi^{\dagger}_{\lambda}(\bm k') H_{\lambda \nu}(\bm k')\psi_{\nu}(\bm k') \right]e^{-i\hat{H}t}\nonumber \\
   &=& e^{i\hat{H} t} \sum_{\bm{k,k'},\lambda,\nu}\left(\hat{\psi}_{\mu}(\bm k)\psi^{\dagger}_{\lambda}(\bm k') \psi_{\nu}(\bm k') -\psi^{\dagger}_{\lambda}(\bm k') \psi_{\nu}(\bm k')\hat{\psi}_{\mu}(\bm k)\right)H_{\lambda \nu}(\bm k')e^{i\bm k \cdot \bm x}e^{-i\hat{H}t}\nonumber \\ &=& e^{i\hat{H} t} \sum_{\bm{k,k'},\lambda,\nu}\left(\hat{\psi}_{\mu}(\bm k)\psi^{\dagger}_{\lambda}(\bm k') \psi_{\nu}(\bm k') +\psi^{\dagger}_{\lambda}(\bm k')\hat{\psi}_{\mu}(\bm k) \psi_{\nu}(\bm k')\right)H_{\lambda \nu}(\bm k')e^{i\bm k \cdot \bm x}e^{-i\hat{H}t}\nonumber \\
   &=& e^{i\hat{H} t} \sum_{\bm{k,k'},\lambda,\nu} \delta_{\bm{kk'}}\delta_{\mu\lambda} \psi_{\nu}(\bm k')H_{\lambda \nu}(\bm k')e^{i\bm k \cdot \bm x}e^{-i\hat{H}t}\nonumber \\
   &=& \sum_{\bm k,\nu} H_{\mu\nu}(\bm k)e^{i\hat{H}t} \hat{\psi}_{\nu}e^{i\bm k\cdot \bm x} e^{-i\hat{H} t}\nonumber \\
   &=& \sum_{\nu} \left(H_{\mu\nu}(-i\bm \nabla) \sum_{\bm k}e^{i\hat{H}t} \hat{\psi}_{\nu}e^{i\bm k \cdot \bm x} e^{-i\hat{H} t}\right)\nonumber \\
   &=& \sum_{\nu} H_{\mu\nu}(-i\bm \nabla) \hat{\psi}_{\nu}(t,\bm x).
   \end{eqnarray}
   Similar, for the creation operator
   \begin{equation}\label{crea-1}
    {\psi}^{\dagger}_{\mu}(t,\bm x)=e^{i\hat{H}t}\hat{\psi}^{\dagger}_{\mu}(\bm x)e^{-i\hat{H}t}=e^{i\hat{H} t}\left(\sum_{\bm k}\hat{\psi}^{\dagger}_{\mu}(\bm k)e^{-i\bm k \cdot \bm x}\right)e^{-i\hat{H}t},
   \end{equation}
   we have
   \begin{eqnarray}\label{Dy-2}
   i\partial_t \hat{\psi}_{\mu}^{\dagger}(t,\bm x)&=&e^{i\hat{H}t}\left[\sum_{\bm k}\hat{\psi}^{\dagger}_{\mu}(\bm k)e^{i\bm k \cdot \bm x},\hat{H} \right] e^{-i\hat{H}t}\nonumber \\
   &=& e^{i\hat{H}t}\sum_{\bm{k,k'},\lambda,\nu}\left[\hat{\psi}^{\dagger}_{\mu}(\bm k)e^{-i\bm k \cdot \bm x},\psi^{\dagger}_{\lambda}(\bm k') H_{\lambda \nu}(\bm k')\psi_{\nu}(\bm k') \right]e^{-i\hat{H}t}\nonumber \\
   &=& e^{i\hat{H} t} \sum_{\bm{k,k'},\lambda,\nu}\left(\hat{\psi}^{\dagger}_{\mu}(\bm k)\psi^{\dagger}_{\lambda}(\bm k') \psi_{\nu}(\bm k') -\psi^{\dagger}_{\lambda}(\bm k') \psi_{\nu}(\bm k')\hat{\psi}^{\dagger}_{\mu}(\bm k)\right)H_{\lambda \nu}(\bm k')e^{-i\bm k \cdot \bm x}e^{-i\hat{H}t}\nonumber \\ &=& e^{i\hat{H} t} \sum_{\bm{k,k'},\lambda,\nu}-\left(\psi^{\dagger}_{\lambda}(\bm k') \hat{\psi}^{\dagger}_{\mu}(\bm k)\psi_{\nu}(\bm k') +\psi^{\dagger}_{\lambda}(\bm k') \psi_{\nu}(\bm k')\hat{\psi}^{\dagger}_{\mu}(\bm k)\right)H_{\lambda \nu}(\bm k')e^{-i\bm k \cdot \bm x}e^{-i\hat{H}t}\nonumber \\
   &=& e^{i\hat{H} t} \sum_{\bm{k,k'},\lambda,\nu} -\delta_{\bm{kk'}}\delta_{\mu\nu} \psi_{\lambda}^{\dagger}(\bm k')H_{\lambda \nu}(\bm k')e^{-i\bm k \cdot \bm x}e^{-i\hat{H}t}\nonumber \\
   &=& -\sum_{\bm k,\lambda} H_{\lambda\mu}(\bm k)e^{i\hat{H}t} \hat{\psi}_{\lambda}^{\dagger}e^{-i\bm k\cdot \bm x} e^{-i\hat{H} t}\nonumber \\
   &=&- \sum_{\lambda} \left(H_{\lambda\mu}(i\bm \nabla) \sum_{\bm k}e^{i\hat{H}t} \hat{\psi}_{\lambda}e^{-i\bm k \cdot \bm x} e^{-i\hat{H} t}\right)\nonumber \\
   &=& -\sum_{\lambda} H_{\lambda\mu}(i\bm \nabla) \hat{\psi}^{\dagger}_{\lambda}(t,\bm x).
   \end{eqnarray}

   The triplet pairs can be described by the G-lessor Green's function $f_{\mu\nu}^<(t,\bm x;t',\bm x')=\langle\hat{\psi}_{\nu}(t',\bm x')\hat{\psi}_{\mu}(t,\bm x)\rangle$ which satisfies
   \begin{eqnarray}\label{Usadel-3}
   i\partial_t f^<_{\mu\nu}&=& \sum_{\lambda}H_{\mu\lambda}(-i\bm \nabla_{\bm x}) f^{<}_{\lambda\nu}(t,\bm x; t',\bm x')\\
   \label{Usadel-3-1}
   i\partial_{t'} f^<_{\mu\nu}&=&\sum_{\lambda}H_{\nu\lambda}(-i\bm \nabla_{\bm x'})f^<_{\mu\lambda}(t,\bm x; t', \bm x').
   \end{eqnarray}
   Because
   \begin{eqnarray}\label{Usadel-4}
   H(-i\bm\nabla)=\left(\begin{array}{cc} -\frac{\nabla^2}{2m}&0 \\ 0& -\frac{\nabla^2}{2m} \end{array}\right)+\bm M \cdot \bm \sigma+\bm h(-i\bm \nabla)\cdot \bm \sigma,
   \end{eqnarray}
   we have
   \begin{eqnarray}\label{Usadel-5}
   H^{\rm T}(-i\bm\nabla)=\left(\begin{array}{cc} -\frac{\nabla^2}{2m}&0 \\ 0& -\frac{\nabla^2}{2m} \end{array}\right)+\bm M \cdot \bm \sigma^*+\bm h(-i\bm \nabla)\cdot \bm \sigma^*.
   \end{eqnarray}
   Combining Eq.~(\ref{Usadel-3},\ref{Usadel-3-1},\ref{Usadel-4},\ref{Usadel-5}), we have
   \begin{eqnarray}\label{Usadel-6}
   i\partial_t f^<(t',\bm x';t,\bm x)&=& H(-i\bm \nabla_{\bm x}) f^{<}(t,\bm x; t',\bm x')\\
   \label{Usadel-6-1}
   i\partial_{t'} f^<(t',\bm x';t,\bm x)&=&f^<_{\mu\lambda}(t,\bm x; t', \bm x')H^{\rm T}(-i\bm \nabla_{\bm x'}).
   \end{eqnarray}
   We define $\bm R=(\bm x+\bm x')/2$, $\bm r=\bm x-\bm x'$, $T=(t+t')/2$, $\tau=t-t'$, $\bm \nabla_{\bm R}=\bm \nabla_{\bm x}+\bm \nabla_{\bm x'}$, $\bm \nabla_{\bm r}=(\bm \nabla_{\bm x}-\bm \nabla_{\bm x'})/2$, $\partial_{T}=\partial_t+\partial_{t'}$ and $\partial_{\tau}=(\partial_t-\partial_{t'})/2$. Therefore, Eq.~(\ref{Usadel-6},\ref{Usadel-6-1}) can be written in the $(T,\tau,\bm R,\bm r)$ coordinates as
   \begin{eqnarray}\label{Usadel-7}
   \rm{Eq}.(\ref{Usadel-6})+\rm{Eq}.(\ref{Usadel-6-1})=i\partial_{T}f^<&=&\left(-\frac{\bm \nabla^2_{\bm r}}{m}-\frac{\bm \nabla^2_{\bm R}}{4m}\right)f^<+\bm M\cdot \bm \sigma f^<+f^< \bm M \cdot \bm \sigma^*+\left(\bm h(-i\bm{\nabla_{R}}/2)\cdot \sigma f^<+f^<\bm h(\bm {-i\nabla_{R}}/2)\cdot\bm\sigma^*\right)\nonumber \\
   &+&\left(\bm{h(-i\nabla_{r})}\cdot \sigma f^<-f^<\bm h(\bm {-i\nabla_{r}})\cdot\bm\sigma^*\right)\\
   \label{Usadel-7-1}
   \rm{Eq}.(\ref{Usadel-6})-\rm{Eq}.(\ref{Usadel-6-1})=2i\partial_{\tau}f^<&=&-\frac{\bm{\nabla_{R}\cdot\nabla_{r}}}{m}f^<+\bm M\cdot \bm \sigma f^<-f^< \bm M \cdot \bm \sigma^*+\left(\bm h(-i\bm{\nabla_{R}}/2)\cdot \sigma f^<-f^<\bm h(\bm {-i\nabla_{R}}/2)\cdot\bm\sigma^*\right)\nonumber \\
   &+&\left(\bm{h(-i\nabla_{r})}\cdot \sigma f^<+f^<\bm h(\bm {-i\nabla_{r}})\cdot\bm\sigma^*\right).
   \end{eqnarray}
   When only impurity scattering is considered, the equation of motion for retarded and G-lesser functions in the center of mass coordinates are the same\cite{SRammer:2007_a}. Therefore, Eq.~\ref{Usadel-7-1} can also be applied for the anomalous retarded Green's function $f^R(E,x)$.
   
   To get a more compact form of Usadel equation for $f^R(E,x)$, we define
   \begin{eqnarray}
   \hat{v}_{\rm so}&=&\frac{\partial \hat{H}_{\rm so}}{\partial \bm k}=(\alpha\sigma_y+\beta \sigma_x)\bm{e_x}-(\alpha \sigma_x+\beta\sigma_y)\bm{e_y},\nonumber \\ \hat{H}_{\rm so}&=&(\beta \hat{p}_x-\alpha \hat{p}_y)\sigma_x+(\alpha \hat{p}_x+\beta \hat{p}_y)=\hat{p}_x(\beta\sigma_x+\alpha\sigma_y)+\hat{p}_y(-\alpha\sigma_x-\beta\sigma_y)=\bm{\hat{p}}\cdot \bm{\hat{v}}_{\rm so},\nonumber \\
   f^R&=&\left(d^R_0\sigma_0+\bm d^R\cdot \bm\sigma\right)i\sigma_y=\mathbb{D}i\sigma_y, \ \ \ \mathbb{D}=d^R_0\sigma_0+\bm d^R\cdot \bm\sigma.
   \end{eqnarray}
   By using the fact that $i\sigma_y \bm \sigma^*=-\bm \sigma i\sigma_y$, the Usadel equation of $f^R$ can be simplified to the equation of $\mathbb{D}$ as
   \begin{eqnarray}\label{Usadel-8}
   i\partial_T \mathbb{D}&=&-\left(\frac{\bm \nabla^2_{\bm r}}{m}+\frac{\bm \nabla^2_{\bm R}}{4m}\right)\mathbb{D}+[(\bm M+\bm h(-i\bm{\nabla_{R}}/2)) \cdot \bm \sigma,\mathbb{D}]+\{\bm h(-i\bm{\nabla_{r}})\cdot \sigma,\mathbb{D}\}, \\
   \label{Usadel-8-1}
   2i\partial_{\tau}\mathbb{D}&=&-i\bm{\nabla_{R}}\{\frac{\hat{\bm v}}{2},\mathbb{D}\}+\{\bm M\cdot \bm \sigma,\mathbb{D}\}+[\bm h(-i\bm{\nabla_{r}})\cdot\bm\sigma,\mathbb{D}].
   \end{eqnarray}
   When $f^R$ slowly varies in the proximity region, it is dominated by Eq.~\ref{Usadel-8-1} which is actually the Elienberger equation in the presence of both magnetization and SOC. Therefore, in this case, it is easily seen that the magnetization will mix the singlet pair with triplet pair which is parallel to the magnetization and the SOC will let the $d$(spin) vector precess in the plane perpendicular to the SOC direction which is similar to the SOC on the spin. When the pair function varies along $x$ direction, the Usadel equation in the presence of only SOC has the form
   \begin{eqnarray}\label{Usadel-1}
   D\nabla^2 d_z&=&4DA d_z-D C\partial_x d_x-D C'\partial_y d_y+2iE\nonumber \\
   D\nabla^2 d_x&=&D (A+B )d_x+D C\partial_x d_z+2iE, \nonumber \\
   D \nabla^2 d_y&=&D (A-B )d_y+D C'\partial_y d_z+2iE,
   \end{eqnarray}
 Here $D=v_f^2\gamma/2$ is the diffusion constant, $\gamma$ is isotropic spin-independent scattering time, $A=2(\alpha^2+\beta^2)m^2$, $B=4\alpha\beta m^2$, $C=4(\alpha+\beta)m$ and $C'=4(\alpha-\beta)m$. In the spin-triplet-superconductor/SOC junction, we assume the \vec{d}-vector of the spin-triplet pairs is along x direction in the bulk of the superconductor. In the SOC region, we assume the $\bm d$-vector only depends on $x$ with the boundary condition ${\bf d}=(d_{0}, 0, 0)$ at the SC/SOC interface and the corresponding solution depends only on $x$, given by
 \begin{eqnarray}\label{Usadel-2}
 \bm d(x,y)=d_{0} e^{-\lambda x}(\cos(qx) \bm{e_x}+\sin(qx)\bm{e_z}).
 \end{eqnarray}
The solution, Eq.(\ref{Usadel-2}), clearly shows the oscillating and decaying behaviors of \vec{d}-vector. When $\alpha=\beta$, we have $4A=A+B=Q^2$ and $C=2Q$. Taking $E=0$, we obtain $\lambda=0$ and $q=Q$, so Eq. (\ref{Usadel-2}) recovers the solution of persistent triplet helix mode in the clean limit (the green lines in Fig.~\ref{shift-trans}). When $\alpha\neq\beta$, the triplet helix mode is no longer conserved. For example, we consider the case with only Rashba SOC and find a damping mode with $q+i\lambda=\sqrt{2\pm 2i\sqrt{7}}Q/4$, as shown in Fig.~\ref{shift-trans}.
\begin{figure}
\centering
\begin{tabular}{l}
\includegraphics[width=0.6\columnwidth]{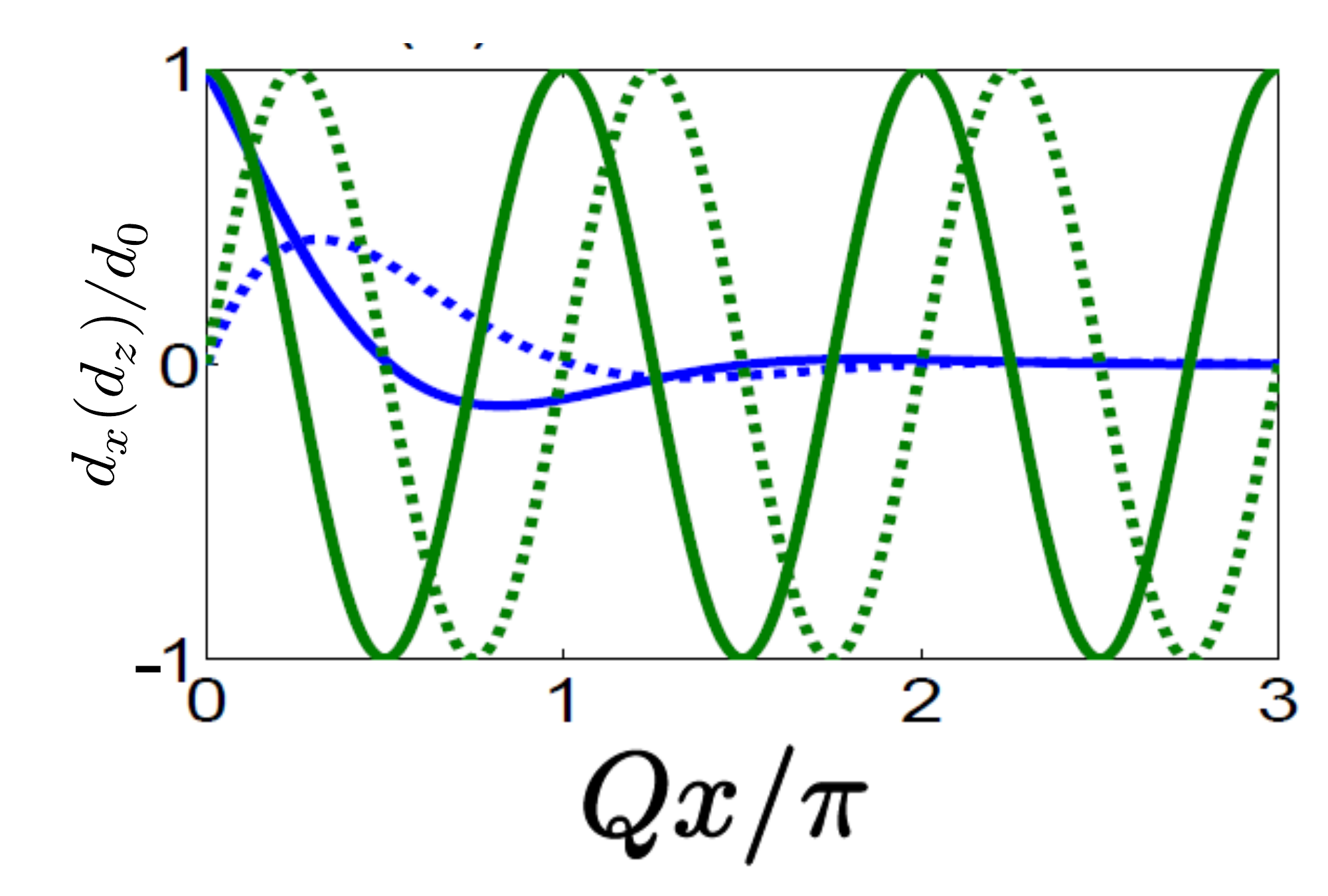}
\end{tabular}
\caption{ The spatial dependence of the ${\bm d}$-vector of triplet pairs for $\alpha=\beta$ (green) and $\alpha=0$ (blue). Green lines ($\alpha=\beta$) show long range oscillations while blues lines ($\alpha=0$) decay rapidly. The solid(dashed) lines indicate $d_x$($d_z$) of triplet pairs. }
\label{shift-trans}
\end{figure}


\end{widetext}

\end{document}